\documentclass[
aip,
apl,
amsmath,amssymb,
reprint,
floatfix]{revtex4-2}
\usepackage{graphicx}
\usepackage{dcolumn}
\usepackage{bm}
\usepackage[utf8]{inputenc}
\usepackage[T1]{fontenc}
\usepackage{mathptmx}
\usepackage{etoolbox}
\usepackage{color}
\usepackage{amsmath, amssymb, array}
\usepackage{booktabs}
\usepackage{makecell}

\usepackage[margin=1in]{geometry}

\usepackage[colorlinks=true, allcolors=blue]{hyperref}

\makeatletter
\def\@email#1#2{%
 \endgroup
 \patchcmd{\titleblock@produce}
  {\frontmatter@RRAPformat}
  {\frontmatter@RRAPformat{\produce@RRAP{*#1\href{mailto:#2}{#2}}}\frontmatter@RRAPformat}
  {}{}
}%
\makeatother
\begin{document}

\preprint{AIP/123-QED}

\title[PHYSICAL REVIEW LETTERS]{Investigation of Laser Plasma Instabilities driven by Coupled High-Power Laser Beams in Magnetized Underdense Plasmas}
\author{C.L.C. Lacoste}
\email{clement.lacoste@u-bordeaux.fr , clement.lacoste@inrs.ca}
\affiliation{CELIA, University of Bordeaux-CNRS-CEA, Talence 33405,  France}
\affiliation{INRS EMT, Varennes J3X 1P7, Canada}

\author{D. Oportus}
\affiliation{Laboratoire pour l’Utilisation des Lasers Intenses, UMR 7605 CNRS-CEA-École Polytechnique- Université Paris VI, 91128 Palaiseau, France}

\author{J. Béard}
\affiliation{CNRS, LNCMI, Univ Toulouse 3, INSA Toulouse, Univ Grenoble Alpes, EMFL, 31400 Toulouse, France}

\author{S.N. Chen}
\affiliation{ELI-NP, ”Horia Hulubei” National Institute for Physics and Nuclear Engineering, 30 Reactorului Street, RO-077125,
Bucharest-Magurele, Romania}

\author{I. Cohen}
\affiliation{Laboratoire pour l’Utilisation des Lasers Intenses, UMR 7605 CNRS-CEA-École Polytechnique- Université Paris VI, 91128 Palaiseau, France}

\author{R. Lelievre}
\affiliation{Laboratoire pour l’Utilisation des Lasers Intenses, UMR 7605 CNRS-CEA-École Polytechnique- Université Paris VI, 91128 Palaiseau, France}

\author{T. Waltenspiel}
\affiliation{CELIA, University of Bordeaux-CNRS-CEA, Talence 33405,  France}
\affiliation{INRS EMT, Varennes J3X 1P7, Canada}
\affiliation{Laboratoire pour l’Utilisation des Lasers Intenses, UMR 7605 CNRS-CEA-École Polytechnique- Université Paris VI, 91128 Palaiseau, France}

\author{W. Yao}
\affiliation{Laboratoire pour l’Utilisation des Lasers Intenses, UMR 7605 CNRS-CEA-École Polytechnique- Université Paris VI, 91128 Palaiseau, France}
\affiliation{Sorbonne Université, Observatoire de Paris, Université PSL, Laboratoire d'étude de l'Univers et des phénomènes eXtrêmes, LUX, CNRS, 75005 Meudon, France}

\author{M. Bardon}
\affiliation{CELIA, University of Bordeaux-CNRS-CEA, Talence 33405,  France}

\author{F.P. Condamine}
\affiliation{The Extreme Light Infrastructure ERIC, ELI Beamines Facility, Za Radnici 835, Doln\`i B\v{r}e\v{z}any 25142, Czech Republic}
\affiliation{GenF, 2 avenue Gay Lussac, 78990 Elancourt, France}

\author{P. Antici}
\affiliation{INRS EMT, Varennes J3X 1P7, Canada}

\author{J. Fuchs}
\affiliation{Laboratoire pour l’Utilisation des Lasers Intenses, UMR 7605 CNRS-CEA-École Polytechnique- Université Paris VI, 91128 Palaiseau, France}
\email{julien.fuchs@polytechnique.edu}

\author{E. D’Humières}
\affiliation{CELIA, University of Bordeaux-CNRS-CEA, Talence 33405,  France}

\date{\today}

\begin{abstract}
Stimulated Brillouin and Raman scattering (SBS and SRS) are instabilities that affect the propagation of high-power lasers  in plasmas. The latter is further affected by Cross-Talk (CT) effects when multiple laser beams are simultaneously propagated in the plasma, as found in the schemes proposed for inertial confinement fusion (ICF). Here we develop a new theoretical model that allows us to evaluate  the impact of CT on SBS and SRS in low-density plasmas. As supported by experiments, we demonstrate that CT  can lead to a reduction of both SBS and SRS, due to the destabilization of the individually triggered instabilities. We further demonstrate that this destabilization effect is accelerated by applying an externally  magnetic field to the plasma, which is also beneficial for the hydrodynamics or fuel heating of ICF. By shedding new light on the promising scheme of magnetized ICF,  our findings thus offer beneficial prospects for 
ICF.


\end{abstract}

\maketitle

In a plasma, which is a non-linear medium \cite{turnbull2017refractive}, a whole range of Laser-Plasma Instabilities (LPI) can be produced. Some of the most well known LPI include Stimulated Raman Scattering (SRS), Stimulated Brillouin Scattering (SBS), but also cross-talk (CT) and cross-beam energy transfer (CBET) between laser beams \cite{nakatsutsumi2010high,michel2010symmetry}. 

In the frame of both indirect-drive Inertial Confinement Fusion \cite{nuckolls1972laser,kruer1991intense,zylstra2022burning} (ICF) and direct-drive ICF \cite{craxton2015direct} it is important to understand these instabilities as they can, through scattering of the laser incident laser, reduce the laser energy that is effectively transferred to the fusion fuel, as well as, through hot electrons generation \cite{Rousseaux1992}, induce undesired fuel pre-heating.
On one hand, LPI processes have the detrimental effect of inducing energy losses in the transport of the laser light before they reach the ICF fuel. On the other hand, LPI can also have beneficial effect in terms of beam smoothing \cite{fuchs2001experimental,loiseau2006laser,maximov2001plasma,malka2003enhanced,depierreux2023experimental,oudin2022cross}, or reducing the growth of subsequent hydrodynamic instabilities
. Complementarily, CT can, through its ability to transfer energy from one laser beam to another via the plasma \cite{nakatsutsumi2010high}, adjust the balance between laser beams that are simultaneously injected in a plasma under different angles, as demonstrated in indirect-drive experiments performed at the National Ignition Facility (NIF) \cite{michel2010symmetry,glenzer2010symmetric}. 
Various strategies for controlling CT, and hence exploit it positively to optimize energy laser deposition on target, have been proposed, including beam diameter reduction \cite{froula2012increasing}, wavelength detuning \cite{marozas2018first}, and laser bandwidth increase \cite{bates2018mitigation}. Note that most of these investigations have been performed in unmagnetized plasma. 


Plasma magnetization has been recently evoked as being able to bring substantial benefits to ICF \cite{moody2022increased,perkins2017potential}, e.g. leading to  reduced hydrodynamic instabilities \cite{walsh2022magnetized}, or to increased fuel temperature \cite{chang2011fusion} and yield \cite{moody2022increased}. However, the detailed effects on the underlying physics still need to be thoroughly assessed to pave the way for potential further improvements. 
Recently, we conducted a study of laser propagation and LPI within a magnetized underdense plasma \cite{yao2023dynamics}, 
and found 
that magnetization 
improved 
the light transmission and 
smoothing of a high-power laser beam propagating through 
a magnetized underdense plasma. Additionally, enhanced SRS backscattering was observed, which is detrimental. This is a phenomenon that our kinetic simulations attributed to the confinement of hot electrons by the externally applied magnetic field \cite{yao2023dynamics}.



In this paper
, we develop a theoretical model to treat the impact of Cross-Talk (CT) onto the growth rate of SBS and SRS. This model is supported by  experimental measurements of SBS and SRS 
within an unmagnetized or magnetized 
plasma, with one or two laser beams, i.e. in conditions where there is no CT, or where CT can be at play. 
In the conditions of our experiment, and consistently with the predictions of the model, we find that CT helps mitigate both SBS and SRS. Indeed, due to the presence of another laser beam, the resonance conditions necessary for the growth of the individual instability driven by one  laser beam cannot be satisfied anymore. This  leads to undermining the  growth of the individual coupling of each beam with the plasma. We further find that the mitigation effect is enhanced by applying the external magnetization onto the plasma. We will first present the experimental setup and results, so that we can contextualize the model we developed to understand the underlying physics.

\begin{figure}[htp]
    \centering
    \includegraphics[width=0.5\textwidth]{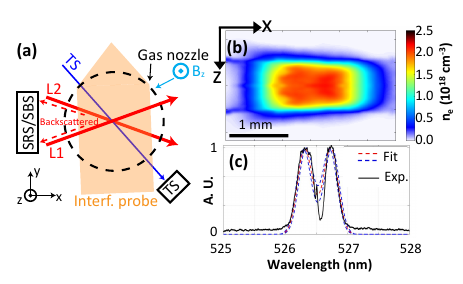}
    \caption{(a) Experimental setup, illustrating the arrangement of laser beams inside the magnetic field coil \cite{yao2023dynamics}, along with the fielded diagnostics (see text). A low-density gas jet is positioned in the center of the coil. The gas flows along the z-axis, and the lasers propagate perpendicularly to the flow. (b) Map of the electron  density in the plasma ionized by the lasers. The image is taken 3 ns after the start of the laser propagation in the gas, and is measured by optical probing (see Supplemental Material \cite{SupplementalMaterial}). The measured plasma density corresponds to the lower end of the plasma densities explored here. The lasers propagates horizontally in the image. (c) Plasma temperature measurement retrieved through Thomson Scattering (TS, see Supplemental Material \cite{SupplementalMaterial}) off the ion waves. The experimental data is shown  as  the black solid line, together with a theoretical fit \cite{doecode_68245} (red dashed lines) and its associated uncertainty bar (blue dashed lines), which gives $T_e = 60 \pm 10$ eV and $T_i = 0.1 T_e$. \textcolor{red}{what is the density for the TS measurement?}
    \label{figure:setup experiment}}
\end{figure}

The experiment was performed at the LULI2000 laser facility (France), see Supplemental Material \cite{SupplementalMaterial} for details. One or two high-power laser beams interacted within a low-density hydrogen gas jet (see Fig. \ref{figure:setup experiment}a), where the gas simulated the
gas fill of indirect-drive ICF hohlraums. 
Both lasers had a wavelength of $\lambda_1$ =1.053~$\mu$m,  a pulse duration of 15~ns, and an on-target intensity of $6 \times 10^{12}$~W/cm$^2$. They were injected in the plasma  with an angle separation of 20°, as used in Inertial Confinement Fusion (ICF) facilities like the NIF.

\begin{figure}[htp]
    \centering
    \includegraphics[width=0.48\textwidth]{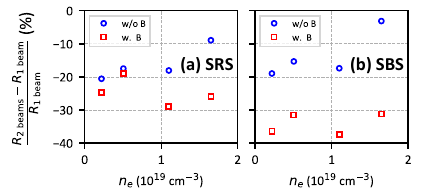}
    \caption{Experimental measurements of the variation of the reflectivities (R) of (a) backward SRS and (b) backward SBS (both normalized to the input laser energy for each shot), as function of the plasma electron density $n_e$, when driving the plasma with  two laser beams, compared to with just one laser beam. 
    The vertical axis is the difference percentage in diode signals between the case of two and one laser beams,  with (red squares) or without (blue circles) magnetic field. The error bars, calculated as the standard deviation of two to three shots per data point, are comparable to the size of the data points.
   }
    \label{figure:diode trends 2}
\end{figure}


The plasma density profile, as measured by our visible optical interferometry (see Fig. \ref{figure:setup experiment}b), had, along the x-axis, a full-width at half-maximum (FWHM) length of 1.5~mm, with the peak electron density being adjustable in the range $n_e~=~
$ $0.2-1.5 \times 10^{19}$~cm$^{-3}$ by modifying the backing pressure of the gas jet system. Note that this range is well within the filling density used in indirect-drive near-vacuum hohlraums \cite{BerzakHopkins2015}, which are used at NIF to optimize laser energy deposition to the hohlraum's walls. \textcolor{red}{Fig.~1c is not referenced anywhere in the text.}

\begin{figure}[htp]
    \centering
    \includegraphics[width=0.48\textwidth]{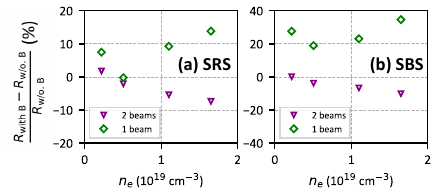}
    \caption{Experimental measurements of the variation of the reflectivities (R) of (a) backward SRS and (b) backward SBS (both normalized to the input laser energy), as function of the plasma electron density $n_e$, when magnetizing the plasma compared to the unmagnetized case. The vertical axis is the difference percentage in diode signals between the case with and without magnetic field, when using two laser beams (purple triangles) or just one beam (green diamonds). The error bars, calculated as the standard deviation of two to three shots per data point, are comparable to the size of the data points.
    }
    \label{figure:diode trends 1}
\end{figure}

 To quantify the level of backward \cite{froula2004full,moody2010backscatter}  Stimulated Brillouin Scattering (SBS) and Stimulated Raman Scattering (SRS) triggered in the plasma, time-resolved diodes were used (see Supplemental Material \cite{SupplementalMaterial}). 
The experimental results are summarized in 
Fig.~\ref{figure:diode trends 2} and Fig.~\ref{figure:diode trends 1}. They both show the difference in the recorded SBS and SRS across a wide range of electron plasma density
. \textcolor{black}{The negative values correspond to a decreased instability and a positive one corresponds to an increased instability.} We compared a number of different cases: either when adding a second laser beam compared to having a single laser beam interacting with the plasma (this is shown in Fig.~\ref{figure:diode trends 2}), or when adding the magnetic field compared to the unmagnetized situation (this is shown in Fig.~\ref{figure:diode trends 1}). 
Fig.~\ref{figure:diode trends 2} clearly demonstrates that
LPI 
is weaker when the two beams are simultaneously injected into the plasma. 



Complementarily, 
Fig.~\ref{figure:diode trends 1} 
demonstrates that, in the case where both laser beams 
are present simultaneously in the plasma, the addition of an external magnetic field results in a decrease of both the SRS and SBS instabilities. This is notably different from the situation where 
only one laser 
beam is 
present, in which case the addition of an external magnetic field results in an increase of both the SRS, consistently with our previous results \cite{yao2023dynamics}, and SBS.

To analyze and interpret the observed trends in the experimental data, we will now detail the analytical model we developed. The model was used to assess the effect of having multiple lasers in the plasma onto the LPIs, as well as to evaluate the effect of the plasma magnetization. The primary objective of this model is to understand the underlying physics behind  the experimentally observed trends of SRS and SBS; however, as detailed below, it is not able to quantitatively assess the magnitude of these instabilities. Various collisionless models of SRS and SBS growth \cite{kruer2019physics}, as well as of the effect of CT, have 
been developed 
\cite{mckinstrie1996two,mckinstrie1998three,kruer1996energy}, in the 
kinetic 
and fluid framework, 
but were mostly restricted to unmagnetized plasmas. C. Greboji et al \cite{grebogi1980brillouin} described  SRS and SBS growth from a single laser beam coupled to an externally magnetized plasma, using fluid assumptions. Their conclusions indicate a reduced SBS and an unchanged SRS, but these were restricted to specific parameters (namely the use of a $CO_2$ laser of wavelength 10.6 $\mu$m), and hence cannot be applied to our conditions. 

\textcolor{black}{As 
detailed below, 
our model 
combines 
fluid and kinetic assumptions. 
The kinetic aspect of the plasma, such as Landau damping, is crucial to have a good understanding of the SRS and SBS behavior. The model considers the general case of the interaction of a single laser beam with a scattered wave, in the presence of a second laser beam, and 
of an external perpendicular magnetic field. 
SRS and SBS are 
considered separately.}

To model SBS, we start by adding a magnetic field in the existing SBS model of C.J. McKinstrie et al. 
\cite{mckinstrie1996two}. In the presence of crossed laser beams, one can describe 
the interaction between two electromagnetic fields in the plasma as follows, where 
Maxwell's wave equation \cite{kruer2019physics,tikhonchuk2023particle} is not modified:

\begin{equation}\label{eq:Mwave_f}
    \left(\partial_t^2 + \omega_{pe}^2 - c^2\nabla^2\right) A= -\omega_{pe}^2n_l A
\end{equation},

and where the ion-acoustic (sound) wave equation becomes: 

\begin{align}\label{eq:force_f}
    \left( \chi \partial_t^2 - c_s^2 \nabla^2 + \gamma \partial_t  \right)n_l  = \frac{c_s^2}{2} \nabla^2 A^2
\end{align}.

with $\chi=Z\left(d\tilde{n_i}/d\tilde{n_e}\right)
=1+k^2 \lambda_D^2$, $\gamma=~\left(Z m_e/m_i\right)\left(1-\chi\right)
\omega_{c} + 2\gamma_s$, $\lambda_D~=\sqrt{\epsilon_0 k_B T_e/n_e}$ is the Debye length, $A$ is the total electromagnetic potential, $\omega_{pe} = \sqrt{\left(n_{e_0} e^2/m_e \epsilon_0\right)}$ 
is the electron plasma frequency, $\omega_c=\left(eB/m_e\right)$
is the cyclotron frequency, B is the external magnetic field, $m_e$ and $m_i$ are respectively the electron and the ion mass, e is the elementary charge, $Z$ the degree of ionization, $n_e=n_{e_0}+\tilde{n_e}$ and $n_i=n_{i_0}+\tilde{n_i}$ are respectively the electron and ion density, $n_l=\tilde{n_e}/n_{e_0}$, $\tilde{n_i}$ and $\tilde{n_e}$ are respectively the low fluctuation of the ion and electron density derived assuming a Boltzmann distribution, as 
described in Ref.\cite{tikhonchuk2023particle}, $n_{e_0}$ and $n_{i_0}$ are respectively the uniform background
electron and ion density, $c_s=\sqrt{Z T_e/m_i}$ is the ion acoustic velocity, $\gamma_s=\sqrt{\frac{\pi}{8}}k v_s \left(\omega_{pi}/\omega_{pe}+c_s^3/v_{Ti}^3 \exp\left(-v_s^2/2v_{Ti}^2\right)\right)$ is the Landau damping frequency due to electrons and ions, $\gamma_{se}=\sqrt{\frac{\pi}{8}}\omega_{pe}/\left(k\lambda_D\right)^3 \exp\left(-3/2-1/2\left(k\lambda_D\right)^2\right)$ is the Landau damping frequency due to electrons \cite{michel2023introduction,tikhonchuk2023particle}, $\omega_{pi}=\sqrt{Z^2e^2n_{i_0}/m_i \epsilon_0}$ is the ion plasma density, $v_s=\sqrt{\frac{ZT_e+3T_i}{m_i}}$ is the sound speed, $v_{Ti}=\sqrt{T_i/m_e}$ is the thermal ion velocity, $k$ is the plasma wave vector, $\epsilon_0$ is the vacuum permittivity and $k_B$ is the Boltzmann constant.

To model SRS, we start by adding the  the magnetic field in  Kruer's model \cite{kruer2019physics}. We conserve 
Maxwell's wave equation (eq. \ref{eq:Mwave_f}) but we replace the ion-acoustic wave by the electron wave equation:

\begin{equation}\label{eq:force_SRS}
    \left(\partial_t^2 + \omega_{pe}^2 - c_e^2 \nabla^2 + \left(\omega_c+2\gamma_{se} \right) \partial_t \right)n_l=  \frac{c_e^2 }{2}\nabla^2 A^2 
\end{equation},

where $c_e=c_s\sqrt{\frac{m_i}{Z m_e}}$ is the electron thermal velocity.
Applying the Fourier transform of these equations and injecting  eq. \ref{eq:force_f} (for SBS) and \ref{eq:force_SRS} (for SRS)  into eq. \ref{eq:Mwave_f}, one can, considering one or two laser beams, compute $\alpha$ and $\beta$, which are respectively the real and imaginary parts of the ion-acoustic or electron response to the ponderomotive force, for SRS and SBS. We define $\alpha_{i-1L}$ and $\beta_{i-1L}$ as the coefficients in the one-beam case and $\alpha_{i-2L}$ and $\beta_{i-2L}$ as those in the two-beam case. Using the slowly varying envelope approximation, such as for one beam, one obtains:

\begin{align} 
    \partial_x A_1=\left(i\alpha_{1-1L} - \beta_{1-1L}\right) A_1 |A_0|^2\\
    \partial_y A_0=\left(i\alpha_{0-1L} + \beta_{0-1L}\right) |A_1|^2 A_0    \label{eq:A_1L_SBS}
\end{align}

with $\omega_{1L}=\omega_1-\omega_0$ and
\[
A = A_1 \exp\left(i\left(k_1 r-\omega_1 t\right)\right) + A_0\exp\left(i\left(k_0 r-\omega_0 t\right)\right) + c.c.
\], $n_l=n (r, t) \times \exp\left[i\left(k_1-k_0\right) r \right] +c.c.$ where $A_1$, $\omega_1$ and $k_1$ are respectively the electromagnetic potential, the frequency and the wave vector of the incident laser beam and $A_0$, $\omega_0$ and $k_0$ are respectively the electromagnetic potential, the frequency and the wave vector of the scattered beam. In the presence of two laser beams, one obtains:

\begin{align} 
    \partial_x A_1=\left(i\alpha_{1-2L} - \beta_{1-2L} \right) A_1 |A_2|^2\\
    \partial_y A_2=\left(i\alpha_{2-2L} + \beta_{2-2L} \right) |A_1|^2 A_2
      \label{eq:A_2L_SBS}
\end{align}

with $\omega_{2L}=\omega_1-\omega_2$ and \[
A = A_1 \exp\left(i\left(k_1 r-\omega_1 t\right)\right) + A_2\exp\left(i\left(k_2 r-\omega_2 t\right)\right) + c.c.
\], $n_l=n (r, t) \times \exp\left[i\left(k_1-k_2\right) r \right] +c.c. $, where $A_2$, $\omega_2$ and $k_2$ are respectively the electromagnetic potential, the frequency and the wave vector of the second laser beam. The 
growth 
of the LPIs in the strong-damping limit is mostly determined by the $\beta$ coefficient \cite{mckinstrie1998three}. To make the problem analytically solvable, we first consider the interaction between the incident laser $A_1$ and the scattered wave $A_0$, followed by the interaction between the two beams $A_1$ and $A_2$. A full four-wave coupling involving $A_0$, $A_1$, $A_2$, and the plasma wave is analytically intractable.
Comparing the ratio between the different values of $\beta$ with and without magnetic field, and for one or two laser beams, we  find the conditions summarized in Table 1 of the Supplemental Material \cite{SupplementalMaterial}.

\begin{figure}[htp]
    \centering
    \includegraphics[width=0.48\textwidth]{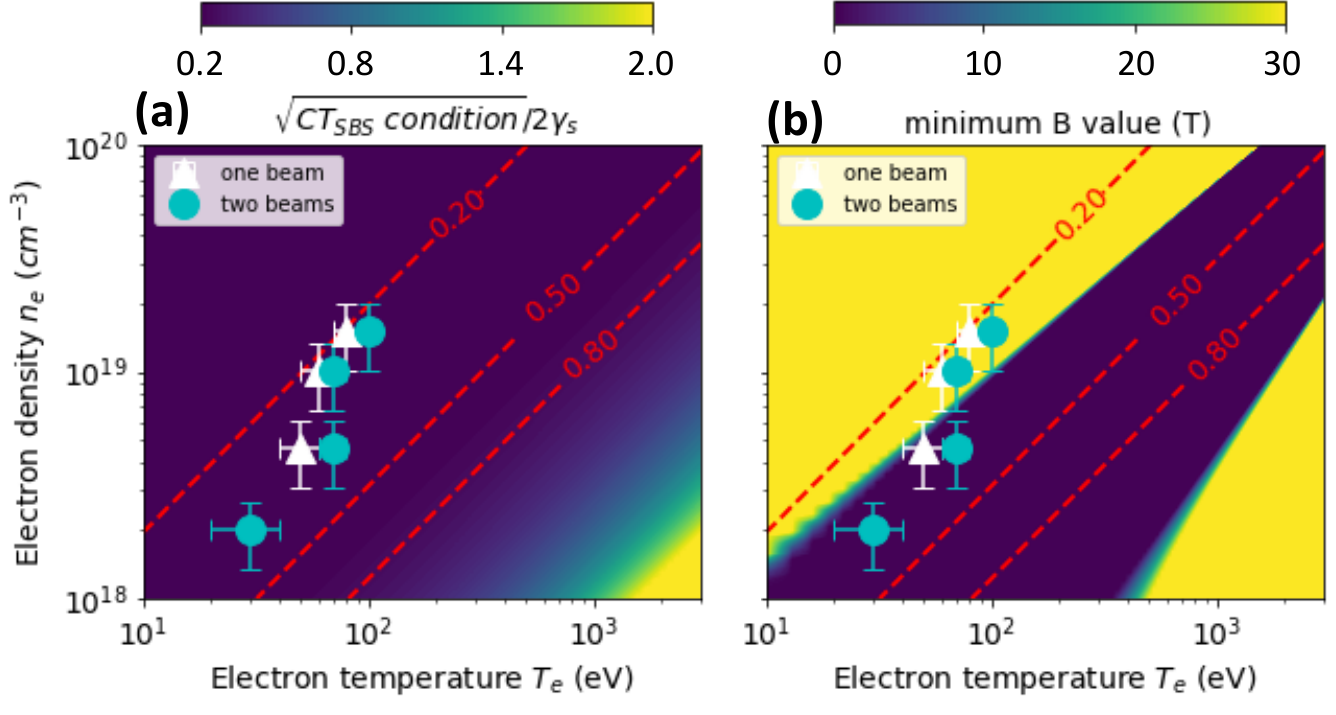}
    \caption{(a) Map of the the square root of the condition for SBS growth, based on our model, in Cross-Talk (CT) conditions, normalized to the Landau damping (as defined in Table S1 of the Supplemental Material \cite{SupplementalMaterial}). 
    (b) Map of the minimum magnetic field required to achieve a decrease single-beam SRS, and correspondingly a two-beam SRS  increase, as a function of the plasma electron density and temperature, when magnetizing the plasma. The red dotted lines correspond to iso-values of the $k\lambda_D$ factor.  The colormap is capped at 30~T to enhance readability. 
    }
    \label{figure:SRS model2}
\end{figure}

We observe that  all the experimental trends that can be inferred from  Fig.~\ref{figure:diode trends 2} and Fig.~\ref{figure:diode trends 1}  are well aligned 
with 
what can be expected from the analytical model. 
Let us first discuss the prediction of the model regarding the impact of CT on SBS in  unmagnetized plasmas. This is illustrated in Fig.\ref{figure:SRS model2}.a, which quantitatively shows the condition for SBS to increase or decrease when using two beams vs one beam, and for various plasma densities and temperatures. When the quantity shown by the color map (which is defined in Table S1 of the Supplemental Material \cite{SupplementalMaterial}) is greater than 1, this corresponds to an expected increase in SBS due to CT. Conversely, when this quantity  is lower than 1, the model predicts that SBS decrease will decrease due to CT. We observe that in the conditions explored in our experiment (represented by the dots in Fig. \ref{figure:SRS model2}.a), we are in a domain where the model predicts that the SBS will always decrease due to CT. This indeed corresponds well to what is experimentally observed, as shown in Fig. \ref{figure:diode trends 2}.b.
 Following Table S1 of the Supplemental Material \cite{SupplementalMaterial}, the same condition can be calculated for SRS in an unmagnetized plasma (not shown here) and the result is the same: a decrease of SRS induced by CT is expected in the conditions of the experiment. This is again well consistent with  the experimental results shown in  Fig.\ref{figure:diode trends 2}.a. Moreover, since, as expected \cite{Akhtar2019,Ivanov2019}, the plasma stays at higher density in the magnetized case (see Fig. S2 of the Supplemental Material \cite{SupplementalMaterial}), this induces a decrease of $k\lambda_D$ and hence of the $\chi$ factor. Note that we did not observe significant changes in the plasma temperature with and without applied magnetic field. The reduction of $\chi$ factor in the magnetized case induces in turn (see  Table S1 of the Supplemental Material \cite{SupplementalMaterial}) that the SBS and SRS are expected to further decrease due to CT in the magnetized case. Again, this is an excellent agreement with what is experimentally observed in Fig.\ref{figure:diode trends 2}.

Fig. \ref{figure:SRS model2}.b further  represents the expected variations in SRS when magnetizing the plasma, compared to the unmagnetized case, as predicted by our model, and corresponding to Table S1 of the Supplemental Material \cite{SupplementalMaterial}. The  color map indicates, for various values of ($n_e,T_e$), the minimum B-field strength, when magnetizing the plasma, for the one beam-induced SRS to increase and for the two-beam SRS to  decrease. One can observe that, except for the lowest plasma density experimentally explored here,  our measurements lie in a region where that minimum required B-field to decrease one beam SRS is high (>30 T). Therefore, given that the magnetic field used in our experiment (20~T) is below the threshold, our model predicts that the applied field enhances the one-beam SRS rather than mitigates it (and vice versa in the two-beam case). This is indeed what we observe in Fig.~\ref{figure:diode trends 1}.a. Note also that for the lowest experimentally investigated density, Fig.~\ref{figure:diode trends 1}.a shows that we observe the inverse behavior, i.e. that of a SRS increase in the two-beam case, which again is well in line with the model, as this point falls in the blue/green region, where our applied 20 T magnetic field is expected to   increase the two-beam SRS. 

As for SBS, based on Table S1 of the Supplemental Material \cite{SupplementalMaterial}, our model predicts that, as the $\chi$ factor is reduced when the plasma is magnetized, this should lead to: (i) a reduction of SBS in the presence of both an external magnetic field and two laser beams (which is mainly due to the reduction of CT by the magnetic field)
, compared to the two-beam case without a magnetic field, and (ii) a  SBS growth  in the presence of an external magnetic field with one beam, compared to the one beam case without magnetic field. Fig.~\ref{figure:diode trends 1}.b shows that these predictions are again in excellent match with the experimental observations.



In summary, we have developed a novel model of SBS and SRS growth in magnetized plasma under the influence of one or two laser beams. Validated by experiments, our model 
shows 
(i) how LPIs growth can be reduced in the presence of coupled laser beams, and (ii) the further benefit brought by the plasma magnetization. Overall, the model predicts that adding a magnetic field influences the SBS by altering the ion-to-electron fluctuation density ratio $\chi$. For SRS, the model predicts that its behaviour 
is strongly tied to $k \lambda_D$ through Landau damping effects. By affecting the plasma conditions ($k \lambda_D$), the magnetic field is able to modify both the SRS and SBS.

The next steps would be to 
systematically investigate 
the effect of the relative orientation of the magnetic field versus the laser propagation axis and polarization, as well as that of varied 
magnetic field strengths. Increasing the plasma temperature to several hundreds of eV  would also allow to raise 
$k \lambda_D$ above 0.4, in which case we should see 
the trends we observe in the present experimental conditions be inverted. For example, we should then observe (i) the SRS increase in the presence of 
two beams and B-field 
and (ii) the SRS, driven by one beam, decrease under the effect of the magnetic field. Note that such latter trends have been observed in previous simulation works \cite{bailly2023validation,winjum2018mitigation}
. The next theoretical developments will focus on establishing an explicit link between $k \lambda_D$ and the plasma magnetization. \\ 



The authors thank the LULI teams for their expert technical support and are grateful to V. Tikhonchuk, M. Bailly-Grandvaux, M. François and D. Viala for their help and useful discussions. This work was supported by the CEA/DAM laser plasma experiments validation project and the CEA/DAM basic technical and scientific studies project. This work was  supported by the National Sciences and Engineering Research Council of Canada (NSERC) (Grant — RGPIN-2023-05459, ALLRP 556340 – 20), Compute Canada (Job: pve-323-ac) as well as the Canada Foundation for Innovation (CFI). This work was supported by the European Research Council (ERC) under the European Union’s Horizon 2020 research and innovation program (Grant Agreement No. 787539; J.F.).  We acknowledge the financial support of the IdEx University of Bordeaux / Grand Research Program "GPR LIGHT", and of the Graduate Program on Light Sciences and Technologies of the University of Bordeaux.

The data that support the findings of this article are not publicly available. The data are available from the authors upon reasonable request.

\bibliographystyle{ieeetr}
\bibliography{biblio.bib}

\end{document}